\documentclass[conference,9pt,dvipsnames]{IEEEtran}
\IEEEoverridecommandlockouts

\AtBeginDocument{%
  \providecommand\BibTeX{{%
    \normalfont B\kern-0.5em{\scshape i\kern-0.25em b}\kern-0.8em\TeX}}}
\usepackage{cite}

\usepackage{url}
\usepackage{enumitem}

\usepackage[utf8]{inputenc} 
\usepackage[T1]{fontenc}    
\usepackage{url}            
\usepackage{booktabs}       
\usepackage{amsfonts}       
\usepackage{amsmath}
\usepackage{nicefrac}       
\usepackage{microtype}      
\usepackage{xcolor}         
\usepackage{colortbl}
\usepackage{graphicx}
\usepackage{svg}
\usepackage{multirow}
\usepackage{wrapfig}
\usepackage{threeparttable}
\usepackage{tikz}
\usepackage{makecell}
\usepackage{tablefootnote}
\usepackage[hidelinks]{hyperref}

\usepackage{color}
\usepackage{xspace}
\usepackage{caption}
\usepackage{adjustbox}

\usepackage{subfigure}
\usepackage{algorithm}
\usepackage{algpseudocode}
\usepackage{fancyhdr}
\usepackage{amsmath}
\usepackage{amssymb}
\usepackage{mathtools}
\usepackage{amsthm}
\usepackage{hyperref}
\usepackage{adjustbox}
\usepackage{xcolor}
\theoremstyle{plain}

\theoremstyle{definition}

\theoremstyle{remark}

\usepackage{tcolorbox}
\tcbuselibrary{listings, breakable}
\usepackage{verbatim}
\usepackage{listings}
\newcommand\BeraMonottfamily{%
  \def\fvm@Scale{0.85}
  \fontfamily{fvm}\selectfont
}
\makeatother
\definecolor{codegreen}{rgb}{0,0.6,0}
\definecolor{codegray}{rgb}{0.5,0.5,0.5}
\definecolor{codepurple}{rgb}{0.58,0,0.82}
\definecolor{backcolour}{rgb}{0.95,0.95,0.95}
\lstdefinestyle{mystyle}{
    backgroundcolor=\color{backcolour},   
    commentstyle=\color{codegreen},
    keywordstyle=\color{magenta},
    numberstyle=\tiny\color{codegray},
    stringstyle=\color{codepurple},
    basicstyle=\BeraMonottfamily\footnotesize,
    breakatwhitespace=false,         
    breaklines=true,                 
    captionpos=b,                    
    keepspaces=true,                 
    numbers=none,                    
    numbersep=5pt,                  
    showspaces=false,                
    showstringspaces=false,
    showtabs=false,                  
    tabsize=2
}
\makeatletter
\long\def\@makefntext#1{\noindent\makebox[0pt][l]{\footnotesize #1}}%
\makeatother

\begin{document}
\pagestyle{plain}
\title{\name: Scaling LLMs with Reasoning Data and Test-Time Compute for Accurate RTL Code Generation
\thanks{\makebox[0pt][l]{979-8-3315-3762-3/25/\$31.00 \copyright~2025 IEEE}}
}


\author{
    {\large Chenhui Deng, Yun-Da Tsai, Guan-Ting Liu, Zhongzhi Yu, Haoxing Ren} \\
    NVIDIA \\
    \{cdeng, yundat, dannliu, zhongzhiy, haoxingr\}@nvidia.com
}


\newcommand{\name}{ScaleRTL\xspace}

\newcommand{\red}[1]{\textcolor{red}{#1}}
\newcommand{\cd}[1]{\textcolor{violet}{[chenhui: #1]}}
\newcommand{\zy}[1]{\textcolor{cyan}{[zhongzhi: #1]}}

\maketitle

\begin{abstract}
Recent advances in large language models (LLMs) have enabled near-human performance on software coding benchmarks, but their effectiveness in RTL code generation remains limited due to the scarcity of high-quality training data. While prior efforts have fine-tuned LLMs for RTL tasks, they do not fundamentally overcome the data bottleneck and lack support for test-time scaling due to their non-reasoning nature. In this work, we introduce ScaleRTL, the first reasoning LLM for RTL coding that scales up both high-quality reasoning data and test-time compute. Specifically, we curate a diverse set of long chain-of-thought reasoning traces averaging 56K tokens each, resulting in a dataset of 3.5B tokens that captures rich RTL knowledge. Fine-tuning a general-purpose reasoning model on this corpus yields ScaleRTL that is capable of deep RTL reasoning. Subsequently, we further enhance the performance of ScaleRTL through a novel test-time scaling strategy that extends the reasoning process via iteratively reflecting on and self-correcting previous reasoning steps. Results show that ScaleRTL achieves state-of-the-art performance on VerilogEval and RTLLM, outperforming 18 competitive baselines by up to $18.4\%$ on VerilogEval and $12.7\%$ on RTLLM.
\end{abstract}

\section{Introduction}
Large language models (LLMs) have progressed from natural‑language understanding to code generation. Frontier models such as GPT~\cite{openai2024gpt4technicalreport}, Gemini~\cite{geminiteam2025geminifamilyhighlycapable}, Claude~\cite{anthropic2023claude}, and DeepSeek~\cite{guo2025deepseek} series achieve near‑human accuracy on software coding benchmarks such as HumanEval~\cite{chen2021humaneval} and MBPP~\cite{austin2021mbpp}. 
This success has sparked interest in applying LLMs to hardware description languages (HDLs), which are central to chip design and verification.
However, due to the low-resource nature of HDLs, there is a scarcity of register-transfer level (RTL) code available for training. As a result, existing general LLMs fall short on RTL benchmarks such as VerilogEval~\cite{liu2023verilogeval} and RTLLM~\cite{lu2024rtllm}. This performance gap underscores the need for domain-adapted LLMs capable of accurate RTL code generation.

Several recent studies have sought to improve LLM performance on RTL coding tasks via supervised fine-tuning, primarily differing in how they construct the training data. CodeV~\cite{zhao2024codev} and MG-Verilog~\cite{zhang2024mg} build high-quality datasets by pairing Verilog modules with multi-level natural language summaries, improving the model’s understanding of RTL semantics. RTLCoder~\cite{liu2024rtlcoder} automates the creation of RTL instruction-code pairs with the aid of GPT-3.5~\cite{openai2023gpt35}. Most recently, CraftRTL~\cite{liu2024craftrtl} addresses key limitations of prior approaches by generating synthetic data that help LLMs understand non-textual representations (e.g., Karnaugh maps) and support targeted code repair, achieving state-of-the-art results on RTL benchmarks.

Despite encouraging progress, RTL-specific LLMs remain fundamentally limited by the scale and quality of their training data. The largest dataset curated in prior work contains 165K samples, each averaging only around 1K tokens, amounting to fewer than 0.2 billion tokens in total~\cite{zhao2024codev}. In contrast, LLM scaling laws~\cite{kaplan2020scaling} suggest that expanding training data to the billion-token scale can yield substantial gains in downstream performance, and RTL coding is unlikely to be an exception. 
Moreover, recent studies have shown that fine-tuning LLMs on high-quality chain-of-thought (CoT) data enables models to perform step-by-step reasoning before generating final answers~\cite{guo2025deepseek}. This CoT fine-tuning substantially improves performance in tasks such as mathematical problem solving~\cite{muennighoff2025s1}. 
However, constructing large-scale RTL reasoning datasets remains an open challenge.

Beyond scaling up training data, a growing body of work highlights the importance of test-time scaling, which extends the reasoning process during model inference. This approach is motivated by recent advances in reasoning LLMs such as OpenAI-o3~\cite{el2025openai_o3} and DeepSeek-R1~\cite{guo2025deepseek}. 
Prior work has shown that longer CoT reasoning traces generally correlate with better model performance in downstream tasks~\cite{kojima2022large, madaan2023selfrefine, muennighoff2025s1, li2025s}. This trend is particularly well-suited to RTL coding, where minor semantic errors or misunderstandings of the design specification can lead to functional failures. By allowing the model to identify mistakes and revise its reasoning before producing a final output, test-time scaling provides a powerful mechanism for improving accuracy in complex hardware design scenarios.

\begin{figure}[t!]
\begin{center}
	\includegraphics[width=\columnwidth]{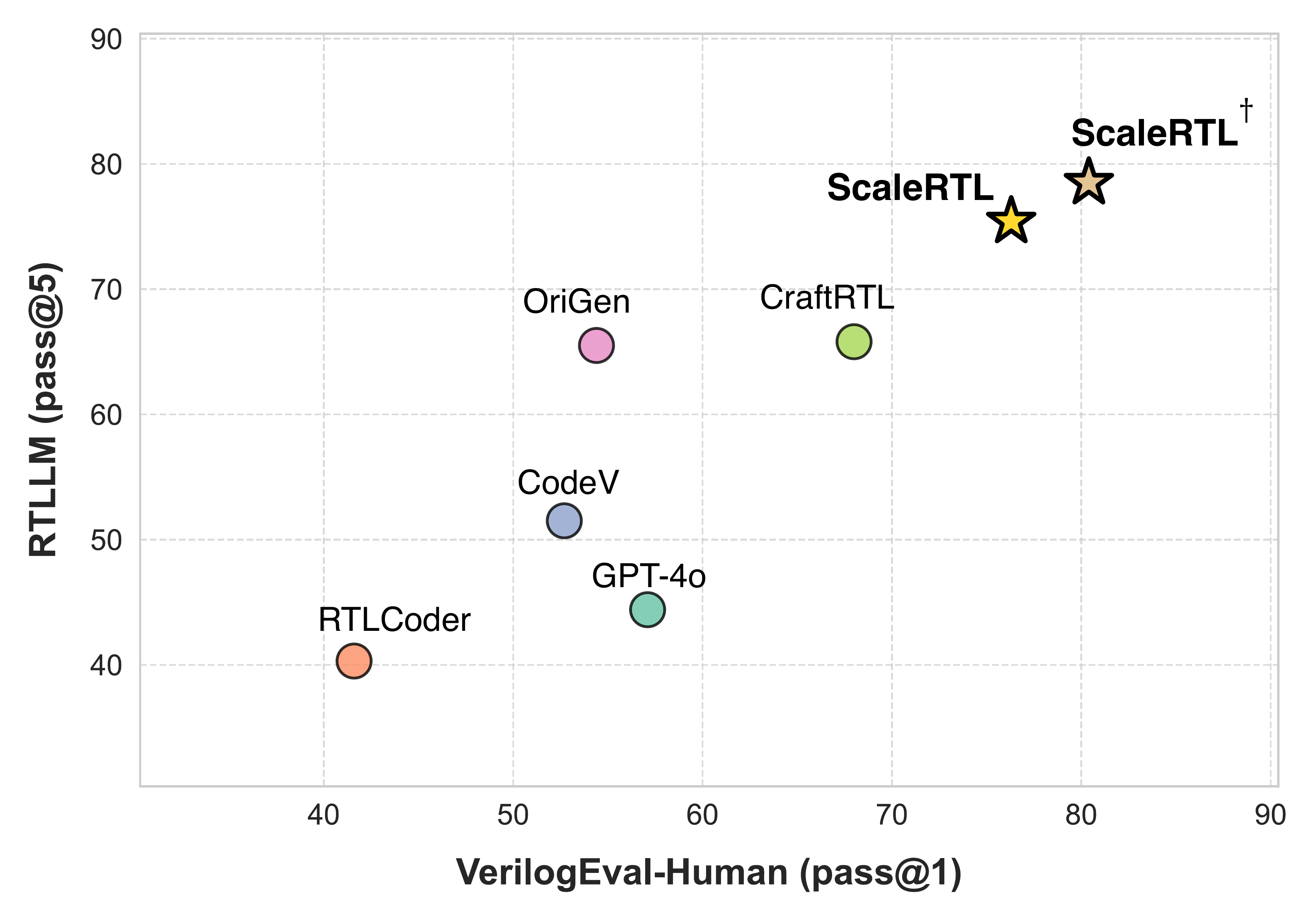}
	\caption{Comparison of LLMs on RTL Coding benchmarks --- \text{\name}$^\dagger$ refers to the test-time scaling variant of \name.} 
     \label{figure:model_compare}
     \vspace{-15pt}
\end{center}
\end{figure}
In this work, we aim to advance the frontier of RTL LLMs through scaling up both training data and test-time reasoning. Concretely, 
we first curate 62K high-quality RTL scripts. Each script is then used to generate a long CoT reasoning trace by prompting DeepSeek-R1, with an average length of 56K tokens. Notably, these long CoT traces significantly enrich the training set despite the limited number of RTL scripts, resulting in a dataset containing 3.5 billion tokens, which is orders of magnitude larger than the data scale considered in previous work. 
Subsequently, we fine-tune the DeepSeek-R1-Distill-Qwen model~\cite{guo2025deepseek} on our curated dataset to obtain \name, a model capable of RTL reasoning. We further scale up test-time compute by iteratively replacing the end-of-reasoning token with a corrective prompt, enabling the model to continuously extend its reasoning traces, yielding \text{\name}$^\dagger$ with improved RTL coding performance.

To demonstrate the efficacy of our approach, we compare \name and \text{\name}$^\dagger$ against $18$ competitive baselines, including 
general-purpose leading LLMs and state-of-the-art RTL models. Experimental results show that our models achieve substantial accuracy improvements over all baselines on two well-established RTL benchmarks: VerilogEval~\cite{liu2023verilogeval} and RTLLM~\cite{lu2024rtllm}.
We summarize our main technical contributions as follows:

\noindent $\bullet$ To our knowledge, this is the first work to develop reasoning LLMs for RTL coding by scaling both reasoning data and test-time compute.

\noindent $\bullet$ To scale up reasoning data for training, we curate a diverse set of long CoT traces, resulting in a 3.5-billion-token corpus. This dataset enables a model to internalize RTL-specific reasoning patterns. 

\noindent $\bullet$ For test-time scaling, we introduce a novel iterative approach that extends the model's reasoning process through rethinking and self-correction, leading to more accurate final solutions.

\noindent $\bullet$ Through large-scale reasoning data and test-time scaling, our approach achieves significant accuracy gains, surpassing previous RTL models by up to $18.4\%$ on VerilogEval and $12.7\%$ on RTLLM.

\section{Background}
\subsection{LLMs for RTL Coding}
\label{llm_rtl}
Adapting large language models (LLMs) to RTL coding has attracted growing interest due to their potential to enhance hardware design productivity. However, general-purpose LLMs exhibit limited performance on RTL benchmarks such as VerilogEval~\cite{liu2023verilogeval}, largely due to the scarcity of domain-specific training data and the absence of dedicated reasoning capabilities. To address this gap, prior work has pursued three major directions: (1) instruction optimization, (2) synthetic data generation, and (3) feedback-driven refinement. In the area of instruction optimization, CodeV~\cite{zhao2024codev} and MG-Verilog~\cite{zhang2024mg} construct high-quality instruction-code datasets by aligning Verilog modules with multi-level natural language summaries, enabling LLMs to better capture RTL semantics across different abstraction levels.
To generate synthetic data, CraftRTL~\cite{liu2024craftrtl} produces diverse, correct-by-construction training samples using artifacts such as Karnaugh maps and state diagrams, while RTLCoder~\cite{liu2024rtlcoder} develops a GPT-assisted pipeline for automated data synthesis and integrates code quality signals during training.
For feedback-driven refinement, OriGen~\cite{cui2024origen} applies compiler-guided self-reflection and code-to-code augmentation to improve functional correctness, and BetterV~\cite{pei2024betterv} employs generative discriminators aligned with synthesis and verification objectives to guide the model toward implementation-consistent outputs.

Despite these advances, most RTL-specialized LLMs remain constrained by limited data scale, as shown in Table \ref{table:dataset}, which limits their ability to handle complex logic and generalize beyond templated patterns. Moreover, these models lack test-time scalability due to their non-reasoning nature, making them incapable of reflection and self-correction to produce accurate solutions.

\begin{table}[t]
\centering
\caption{Dataset statistics for various RTL LLM methods --- $^*$ indicates estimated values based on maximum sequence length.}
\label{table:dataset}
\begin{adjustbox}{width=\columnwidth,center}
\begin{tabular}{lcccc}
\toprule
& \textbf{\# Samples} & \textbf{Avg. Tokens per Sample} & \textbf{Total Tokens} \\
\midrule
RTLCoder~\cite{liu2024rtlcoder} & 26K & 0.2K & 6M \\
MG-Verilog~\cite{zhang2024mg} & 11K & 1.5K & 16M \\
OriGen~\cite{cui2024origen} & 222K & 0.6K & 139M \\
CodeV~\cite{zhao2024codev} & 165K & 1K$^*$ & 165M$^*$ \\
CraftRTL~\cite{liu2024craftrtl} & 110K & 1K & 110M \\
\midrule
ScaleRTL (Ours) & 62K & 56K & 3.5B \\
\bottomrule
\end{tabular}
\end{adjustbox}
\end{table}

\subsection{Scaling Laws for LLMs}
\label{scale_law}
Scaling laws have been widely recognized as empirical principles for improving LLM performance across diverse domains. They characterize how model capability on downstream tasks improves with increased computation, both during training and inference. On the one hand, at training time, foundational studies observe that LLMs' performance improves predictably with scale, following a power-law relationship with respect to the logarithm of training compute~\cite{kaplan2020scaling, hoffmann2022training}. 
On the other hand, test-time scaling has emerged as a complementary strategy for enhancing model performance. Rather than increasing model capacity or expanding the training dataset, test-time scaling focuses on allocating more computational resources during inference. This includes enabling longer CoT reasoning~\cite{wei2022chain, kojima2022large} or iterative decoding~\cite{chen2023replug, madaan2023selfrefine} to extend the model’s reasoning horizon and improve output quality.
Recent studies have shown that sequential test-time scaling, where the model extends its reasoning process to rethink and self-correct its own steps, can lead to substantial gains on mathematical~\cite{muennighoff2025s1} and software coding tasks~\cite{li2025s}.
These insights highlight the value of scaling along both axes, motivating our work to jointly explore training-time and test-time scaling for advancing LLMs in RTL code generation.

Notably, while agentic methods can also be broadly categorized under test-time scaling, they typically focus on orchestrating multiple tool-augmented actions or planning steps. This is orthogonal to our work, which exclusively focuses on scaling test-time compute based on the intrinsic capability of a single reasoning LLM.



\section{Methodology}
\label{method}
\begin{figure*}[t!]
\begin{center}
	\includegraphics[width=\textwidth]{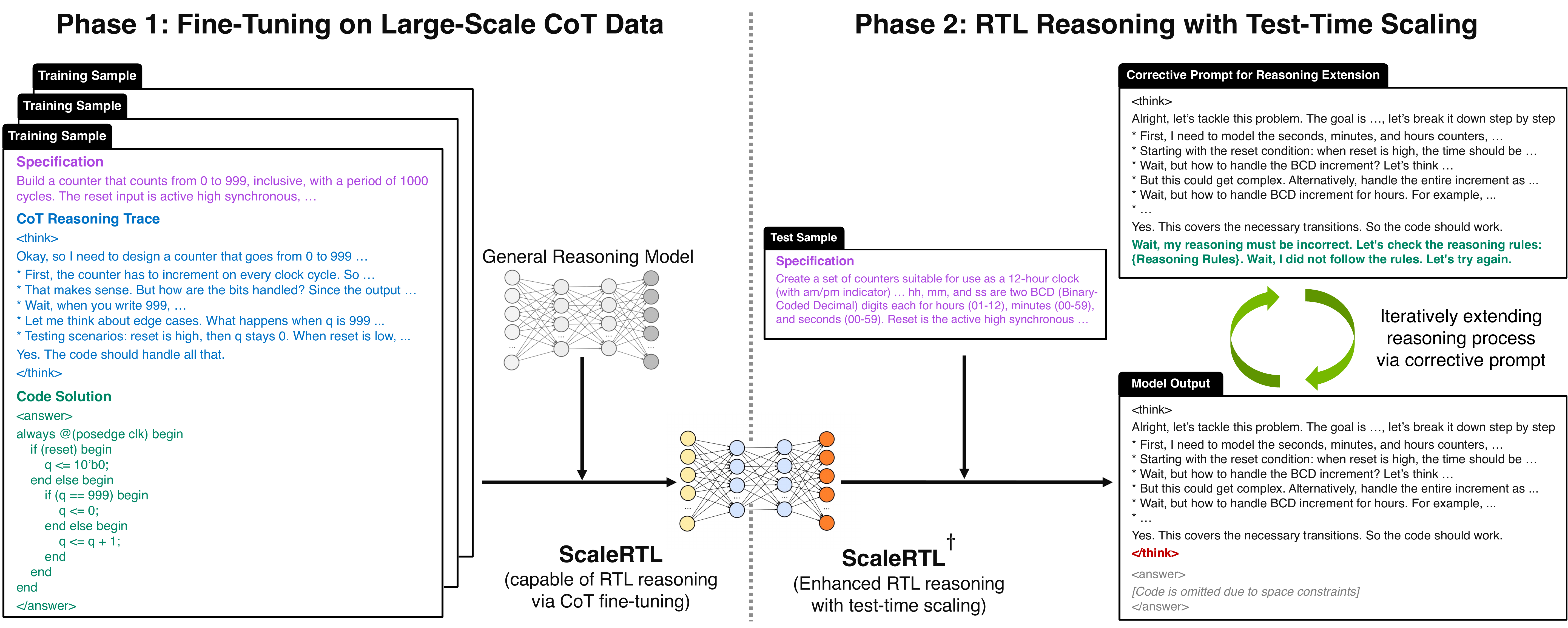}
	\caption{An overview of our approach, which consists of two phases --- In the first phase, we fine-tune a model on a large-scale curated CoT reasoning dataset to obtain \name. In the second phase, we apply test-time scaling to \name to further enhance its RTL reasoning capability, resulting in \text{\name}$^\dagger$.} 
     \label{figure:overview}
     \vspace{-15pt}
\end{center}
\end{figure*}
Figure \ref{figure:overview} presents an overview of our approach, which comprises two main phases. The first phase curates a billion-token scale CoT dataset for fine-tuning, resulting in the \name model that is capable of reasoning through RTL problems before generating final solutions. The second phase introduces a novel iterative approach \text{\name}$^\dagger$, which further enhances RTL reasoning by enforcing the model to rethink and self-correct its previous steps in the reasoning trace. In the following sections, we detail our data curation process and model fine-tuning in Section~\ref{cot_train}, and then describe our test-time scaling strategy in Section~\ref{test_scale}.

\subsection{Large-Scale RTL Data Curation for Fine-Tuning}
\label{cot_train}
\paragraph{\textbf{High-Quality RTL Script Collection}}
We begin by collecting 5 million raw RTL code snippets scraped from public web sources, comprising over 189 billion tokens. To ensure data quality and diversity, we apply a multi-stage filtering pipeline. First, we perform data deduplication to remove identical files. We then apply rule-based filtering to eliminate extremely long scripts (exceeding 64K tokens), which are typically machine-generated and tend to be of lower quality. Next, we perform syntax validation on the remaining scripts to ensure correctness.

To prevent data contamination of downstream RTL benchmarks, we compute the Jaccard similarity between each collected code sample and the golden solutions in the benchmarks. Specifically, we represent each script as a set of unique 5-grams (i.e., consecutive token sequences of length 5), and compute similarity as follows. Given a collected script \( A \) and a golden script \( B \), their Jaccard similarity is defined as:
\[
\text{Jaccard}(A, B) = \frac{|T_5(A) \cap T_5(B)|}{|T_5(A) \cup T_5(B)|}
\]
where \( T_5(A) \) and \( T_5(B) \) denote the sets of unique 5-grams extracted from scripts \( A \) and \( B \), respectively.
We discard any script whose similarity score exceeds $0.8$ with respect to any golden reference. 

In addition, we apply an embedding-based semantic similarity filter~\cite{hui2024qwen2} to discard code snippets that deviate significantly from the distribution of RTL benchmarks. This step helps eliminate outliers examples from the dataset.
After all filtering stages, we retain 62K high-quality RTL scripts containing approximately 200 million tokens, which serve as the basis for our subsequent CoT data generation.

\paragraph{\textbf{Billion-Token Scale CoT Data Generation}}
To substantially expand the scale of our dataset, we generate a long CoT reasoning trace for each of the 62K curated high-quality RTL scripts, using DeepSeek-R1~\cite{guo2025deepseek}, a state-of-the-art general-purpose reasoning LLM.
We generate the CoT data in two steps. First, following the \textsc{OSS-Instruct} data generation strategy~\cite{wei2023magicoder}, we prompt DeepSeek-R1 to formulate a coding problem (i.e., a specification) based on each RTL script. After obtaining 62K specifications from our curated RTL corpus, we prompt DeepSeek-R1 again, using these specifications as input, to generate detailed CoT reasoning traces along with the corresponding code solutions. Our detailed prompts for both steps are available in Appendix \ref{prompt_spec}.

It is worth noting that DeepSeek-R1 natively wraps its CoT reasoning traces within \texttt{<think>...</think>} tags and its final solutions within \texttt{<answer>...</answer>} tags. This structured formatting allows us to directly and reliably extract the reasoning trace and final output as separate components.
This process yields high-quality <specification, CoT reasoning trace, code solution> triples that form our dataset. In total, we generate 62K CoT-annotated samples, each averaging 56K tokens, resulting in a corpus of approximately 3.5 billion tokens. Notably, this dataset is orders of magnitude larger than those used in prior work, as summarized in Table \ref{table:dataset}.
We ensure that our CoT dataset contains no specification or code overlapping with existing RTL benchmarks, thereby avoiding any data contamination.


\paragraph{\textbf{Model Fine-Tuning}} After constructing the CoT dataset, we perform supervised fine-tuning (SFT) on the general reasoning model DeepSeek-R1-Distill-Qwen, resulting in a model specifically tailored for RTL reasoning. 
Concretely, each training example is represented as a triplet 
$\langle x, r, y \rangle$, where $x$ denotes the specification, $r$ is the CoT reasoning trace, and $y$ is the final RTL code solution.
Let $\mathcal{D} = \{(x^{(i)}, r^{(i)}, y^{(i)})\}_{i=1}^N$ be the fine-tuning dataset, where $N$ is the total number of training samples.
We optimize the model to maximize the conditional likelihood of both the reasoning trace and the code output, given the specification:
\begin{align*}
\mathcal{L}_{\text{SFT}}(\theta) = - \sum_{i=1}^N \Bigg[ 
& \sum_{t=1}^{|r^{(i)}|} \log P_\theta\left(r^{(i)}_t \mid x^{(i)}, r^{(i)}_{<t} \right) \\
+ & \sum_{t=1}^{|y^{(i)}|} \log P_\theta\left(y^{(i)}_t \mid x^{(i)}, r^{(i)}, y^{(i)}_{<t} \right) 
\Bigg]
\end{align*}
This formulation encourages the model to first internalize the reasoning steps and subsequently generate the correct RTL implementation, thus promoting better generalization and interpretability in RTL coding.

\subsection{RTL Reasoning Enhancement via Test-Time Scaling}
\label{test_scale}
After fine-tuning on the CoT data, we obtain a model capable of generating RTL-specific reasoning traces before producing a final solution. However, the model may still follow flawed reasoning paths, leading to solutions that fail to pass the testbench. Motivated by recent advances in test-time scaling for mathematical problem solving, we aim to extend the model's reasoning trace at inference time, enabling it to rethink and self-correct earlier steps.

As introduced in s1~\cite{muennighoff2025s1}, a straightforward approach to test-time scaling is to append a ``\texttt{Wait}'' token at the end of the reasoning trace to prompt the model to continue its reasoning process. While prior work has shown that this method effectively lengthens reasoning traces and improves accuracy on mathematical benchmarks, we observe that it fails to elicit deeper reasoning from our RTL model. Instead, the model tends to agree with its prior conclusion without revisiting earlier logic. As a result, this simple trick often reproduces the original code solution generated by \name and fails to correct any errors.

\begin{algorithm}[t]
\caption{Iterative Test-Time Scaling for RTL Coding}
\label{alg:rtl_scaling}
\begin{algorithmic}[1]
\State \textbf{Input:} Fine-tuned model $M$, test problem $P$, reasoning rules $R$, maximum iterations $T$
\State \textbf{Output:} Final reasoning trace $\mathcal{T}$ and RTL solution $S$

\State Initialize reasoning trace $\mathcal{T} \gets \text{Inference}(M, P)$
\State Extract initial RTL solution $S \gets \text{ExtractSolution}(\mathcal{T})$
\State $t \gets 0$
\While{$t < T$ \textbf{and} \textbf{not} \text{IsCorrect}($S$)}
    \State Remove RTL solution from $\mathcal{T}$ after delimiter ``\texttt{</think>}'' 
    \State Append corrective prompt:
    \State \hskip1em ``\textit{Wait, my reasoning must be incorrect. Let's check the reasoning rules: \{$R$\}. Wait, I did not follow the rules. Let's try again.}''
    \State $\mathcal{T} \gets \text{Inference}(M, \mathcal{T})$
    \State $S \gets \text{ExtractSolution}(\mathcal{T})$
    \State $t \gets t + 1$
\EndWhile

\State \Return $\mathcal{T}, S$
\end{algorithmic}
\end{algorithm}
Algorithm~\ref{alg:rtl_scaling} presents the complete procedure of our novel test-time scaling method for RTL coding.
For each problem in the downstream benchmarks, we first perform standard inference using the fine-tuned model to generate a reasoning trace and the corresponding RTL solution. If the solution fails to pass the testbench, the end-of-reasoning token delimiter ``\texttt{</think>}'' is then replaced with a corrective prompt:
``\textit{Wait, my reasoning must be incorrect. Let's check the reasoning rules: \{Reasoning Rules\}. Wait, I did not follow the rules. Let's try again.}''
The \textit{\{Reasoning Rules\}} are generated offline and consist of general RTL coding guidelines, such as \textit{``Carefully implement output signals based on their timing requirements''}. We provide details on generating such rules in the Appendix \ref{prompt_rule}.
To enable rethinking, we discard the previously generated RTL code solution following ``\texttt{</think>}'' and append the corrective prompt to the remaining reasoning trace. The modified trace is then fed back into the model to continue the reasoning and produce a revised RTL solution.
We find that this encourages the model to reflect on and self-correct earlier mistakes, often resulting in a correct solution. This process can be applied iteratively by repeatedly replacing ``\texttt{</think>}'' with the corrective prompt to scale up test-time reasoning. 
As a result, it significantly increases the number of reasoning tokens at inference time and leads to improved RTL coding performance.

Notably, while several agent-based approaches for RTL coding (e.g., VerilogCoder~\cite{ho2025verilogcoder}) can be broadly viewed as test-time scaling methods by iteratively debugging code that fails the testbench, they rely on general-purpose LLMs (e.g., GPT-4o) and primarily focus on designing task planners and integrating external tools (e.g., waveform tracing). In contrast, we take an orthogonal approach by leveraging the intrinsic reasoning capabilities of \name to extend reasoning traces and automatically correct earlier reasoning errors. Integrating \name with existing agentic approaches is left for future work.

\section{Experiment}
We conduct a comprehensive evaluation of \name and its test-time scaling variant \text{\name}$^\dagger$, comparing their performance against 18 competitive LLMs, including 7 state-of-the-art RTL coding baselines. Moreover, we perform a detailed analysis of the impact of test-time scaling. Finally, to ensure that \name does not overfit to RTL benchmarks, we evaluate it on general-purpose coding benchmarks, including HumanEval, MBPP (Mostly Basic Python Programming), and IFEval for instruction-following evaluation.

\subsection{Experimental Setup}
\label{setup}
\paragraph{\textbf{Baselines}} 
We compare \name against a diverse set of LLMs, spanning general-purpose foundation and coding models to RTL-specialized models. The foundation model category includes LLaMA3.1-405B~\cite{grattafiori2024llama}, GPT-4o~\cite{hurst2024gpt}, and Nemotron4-340B~\cite{adler2024nemotron}. General coding models, including CodeLlama~\cite{roziere2023code}, Starcoder2~\cite{lozhkov2024starcoder}, and DeepSeek-Coder~\cite{zhu2024deepseek}, are trained on broad programming corpora but lack RTL-specific adaptation. RTL coding models encompass state-of-the-art methods such as CodeV, CraftRTL, and RTLCoder, which are fine-tuned to generate Verilog from natural language descriptions. 
To eliminate the potential advantage of larger model size in \name and better attribute performance gains to reasoning ability, we fine-tune LLaMA3.3-70B~\cite{grattafiori2024llama} on the CraftRTL dataset, denoted as CraftRTL-70B. This serves as a strong baseline for isolating the effect of model scale.
Lastly, we evaluate general reasoning models such as R1-Distill-Qwen-32B and R1-Distill-Llama-70B, which are trained for CoT reasoning but are not explicitly designed for RTL tasks.

For more recent foundation models such as DeepSeek-R1 and Claude3.7-Sonnet, we intentionally exclude them from our main comparisons, as their training data cutoff dates postdate the release of RTL benchmarks. It is likely that these models have been exposed to benchmark data during pre-training. For completeness, we provide comparative results with these models in the Appendix \ref{sota_llms}.

\paragraph{\textbf{Dataset and Model Training}} 
We obtain the \name model by fine-tuning DeepSeek-R1-Distill-Qwen-32B on our curated CoT dataset containing 3.5B tokens shown in Table \ref{table:dataset}. We use a learning rate of $0.0001$, a global batch size of $128$, $5$ training epochs, and $64$ warmup steps. Our model is trained on $32$ nodes, each equipped with $8$ A100 GPUs. To accommodate long CoT sequences, we pack the training data into chunks of $32,768$ tokens and use a context parallelism of $4$ and a tensor parallelism of $4$.

\paragraph{\textbf{Model Inference}}
We use vLLM~\cite{kwon2023efficient} with \texttt{bf16} precision, a tensor parallelism factor of $4$, a temperature of $0.2$, and an initial token limit of $16,384$. For \text{\name}$^\dagger$, we apply two additional inference passes per test problem using our test-time scaling strategy, as discussed in Section~\ref{test_scale}. The first pass uses a token limit of $32,768$, and the second extends it to $49,152$.

\paragraph{\textbf{RTL Benchmarks and Evaluation Metrics}}
We evaluate model performance on two representative RTL code generation benchmarks: VerilogEval~\cite{liu2023verilogeval} and RTLLM~\cite{lu2024rtllm}, which are designed to assess RTL code generation from natural language specifications. Notably, VerilogEval consists of two subsets: VerilogEval-Human, which contains $156$ manually adapted problem descriptions from the HDLBits platform, and VerilogEval-Machine, which includes $143$ descriptions automatically generated by GPT-3.5. We focus on measuring functional correctness in these benchmarks, as it provides a more practical assessment than syntax correctness.

Following prior work~\cite{liu2023verilogeval}, we use the unbiased \textit{pass@k} metric to evaluate functional correctness. This metric estimates the probability that at least one of \(k\) generated outputs passes all test cases, given \(n \geq k\) total attempts per problem. We choose $n=10$ in this work.

\begin{table*}[t!]
\centering
\caption{Functional Correctness on RTL benchmarks --- We highlight the top \textcolor{Green}{\textbf{first}}, \textcolor{NavyBlue}{\textbf{second}}, and \textcolor{Purple}{\textbf{third}} results per benchmark.}
\label{table:rtl}
\begin{tabular}{clcccccc}
\toprule
\multirow{2}{*}{\textbf{Type}} & \multirow{2}{*}{\textbf{Model}} &
\multicolumn{2}{c}{\textbf{VerilogEval-Machine}} & \multicolumn{2}{c}{\textbf{VerilogEval-Human}} & \multicolumn{2}{c}{\textbf{RTLLM v1.1}} \\
& & \textbf{pass@1} & \textbf{pass@5} & \textbf{pass@1} & \textbf{pass@5} & \textbf{pass@1} & \textbf{pass@5} \\
\midrule
\multirow{4}{*}{Foundation Models} 
& Llama3.1-405B & 67.3 & 75.1 & 53.8 & 61.0 & 38.9 & 45.8 \\
& Llama3.3-70B & 63.9 & 67.3 & 46.5 & 49.3 & 12.7 & 22.5 \\
& Nemotron4-340B & 53.0 & 60.3 & 43.1 & 48.3 & 18.9 & 20.7 \\
& GPT-4o & 65.9 & 71.4 & 57.1 & 63.9 & 33.8 & 44.4 \\
\midrule
\multirow{5}{*}{General Coding Models} 
& CodeLlama-7B & 43.1 & 47.1 & 18.2 & 22.7 & 17.9 & 29.9 \\
& CodeQwen-7B & 46.5 & 54.9 & 22.5 & 26.1 & 24.1 & 34.0 \\
& Starcoder2-15B & 68.7 & 82.3 & 37.7 & 50.6 & 15.5 & 37.6 \\
& DeepSeek-Coder-V2-16B & 67.4 & 78.3 & 46.9 & 55.9 & 33.1 & 37.1 \\
& DeepSeek-Coder-V2-236B & 68.2 & 74.1 & 56.4 & 62.2 & 34.5 & 50.2 \\
\midrule
\multirow{7}{*}{RTL Coding Models}
& RTLCoder-6.7B & 61.2 & 76.5 & 41.6 & 50.1 & 35.8 & 40.3 \\
& BetterV-6.7B & 67.8 & 79.1 & 45.9 & 53.3 & $-$ & $-$ \\
& CodeV-6.7B & 77.9 & \textcolor{NavyBlue}{\textbf{88.6}} & 52.7 & 62.5 & 42.4 & 51.5 \\
& OriGen-6.7B & 74.1 & 82.4 & 54.4 & 60.1 & $-$ & 65.5 \\
& MG-Verilog-7B & 52.7 & 58.5 & $-$ & $-$ & $-$ & $-$ \\
& CraftRTL-15B & \textcolor{NavyBlue}{\textbf{81.9}} & 86.9 & 68.0 & \textcolor{Purple}{\textbf{72.4}} & \textcolor{Purple}{\textbf{49.0}} & \textcolor{Purple}{\textbf{65.8}} \\
& CraftRTL-70B & \textcolor{Purple}{\textbf{80.1}} & 81.9 & \textcolor{Purple}{\textbf{69.7}} & 71.5 & 46.0 & 52.7 \\
\midrule
\multirow{2}{*}{General Reasoning Models}
& R1-Distill-Qwen-32B & 55.2 & 64.3 & 45.5 & 66.0 & 35.8 & 51.7 \\
& R1-Distill-Llama-70B & 62.7 & 66.9 & 52.5 & 71.8 & 15.2 & 34.4 \\
\midrule
\multirow{2}{*}{RTL Reasoning Models}
& ScaleRTL-32B & 80.0 & \textcolor{Purple}{\textbf{88.3}} & \textcolor{NavyBlue}{\textbf{76.3}} & \textcolor{NavyBlue}{\textbf{88.4}} & \textcolor{NavyBlue}{\textbf{55.5}} & \textcolor{NavyBlue}{\textbf{75.4}} \\
& \text{\name}$^\dagger$-32B & \textcolor{Green}{\textbf{83.2}} & \textcolor{Green}{\textbf{89.1}} & \textcolor{Green}{\textbf{80.4}} & \textcolor{Green}{\textbf{90.8}} & \textcolor{Green}{\textbf{60.7}} & \textcolor{Green}{\textbf{78.5}} \\
\bottomrule
\vspace{-15pt}
\end{tabular}
\end{table*}
\subsection{Main Results}
\label{rtl}
Table~\ref{table:rtl} clearly demonstrates the effectiveness of our proposed model, \name, and its test-time scaling variant, \text{\name}$^\dagger$. Specifically, our best model improves pass@1 over the strongest prior RTL-specific baseline by up to $10.7\%$ on VerilogEval and $11.7\%$ on RTLLM. The improvements are also significant in pass@5, with gains of $18.4\%$ on VerilogEval and $12.7\%$ on RTLLM. These results highlight the importance of incorporating domain-specific reasoning traces, rather than relying solely on instruction-code fine-tuning.

When comparing \name to general reasoning models such as R1-Distill-Qwen and R1-Distill-Llama, the advantage becomes even more pronounced. Despite these baselines being optimized for CoT reasoning, they fall short in RTL tasks. This suggests that generic reasoning ability is not sufficient; accurate RTL coding requires reasoning that is explicitly aligned with hardware design knowledge.

Moreover, the performance improvement of \name does not simply stem from increased model size. For instance, although CraftRTL-70B is significantly larger, it still underperforms compared to our 32B \name model across all benchmarks. This underscores that scaling model parameters alone is insufficient; fine-tuning on RTL reasoning traces is essential for accurate RTL code generation.

Finally, the performance gap between \name and \text{\name}$^\dagger$ confirms that test-time scaling plays a critical in accurate RTL code generation. By allowing the model to reflect and revise its outputs over multiple inference passes, we observe consistent improvements across all benchmarks. This validates test-time scaling as an effective mechanism for enhancing model accuracy without retraining. We provide a deeper analysis of test-time scaling in Section~\ref{test_scale_analysis}. 
In addition to the standard pass@k metric used for all baselines, we also report the pass rate, which measures whether a problem can be solved at least once within $n=10$ trials. We believe this is a more meaningful indicator of the effectiveness of our test-time scaling approach, as it directly reflects the model's ability to eventually arrive at a correct solution through iterative reasoning. The pass rate results for both \name and \text{\name}$^\dagger$ are provided in Appendix \ref{pass_rate}.

\subsection{Test-Time Scaling Analysis}
\label{test_scale_analysis}
\begin{figure*}[ht]
    \centering
    \subfigure[]{%
        \includegraphics[width=0.32\textwidth]{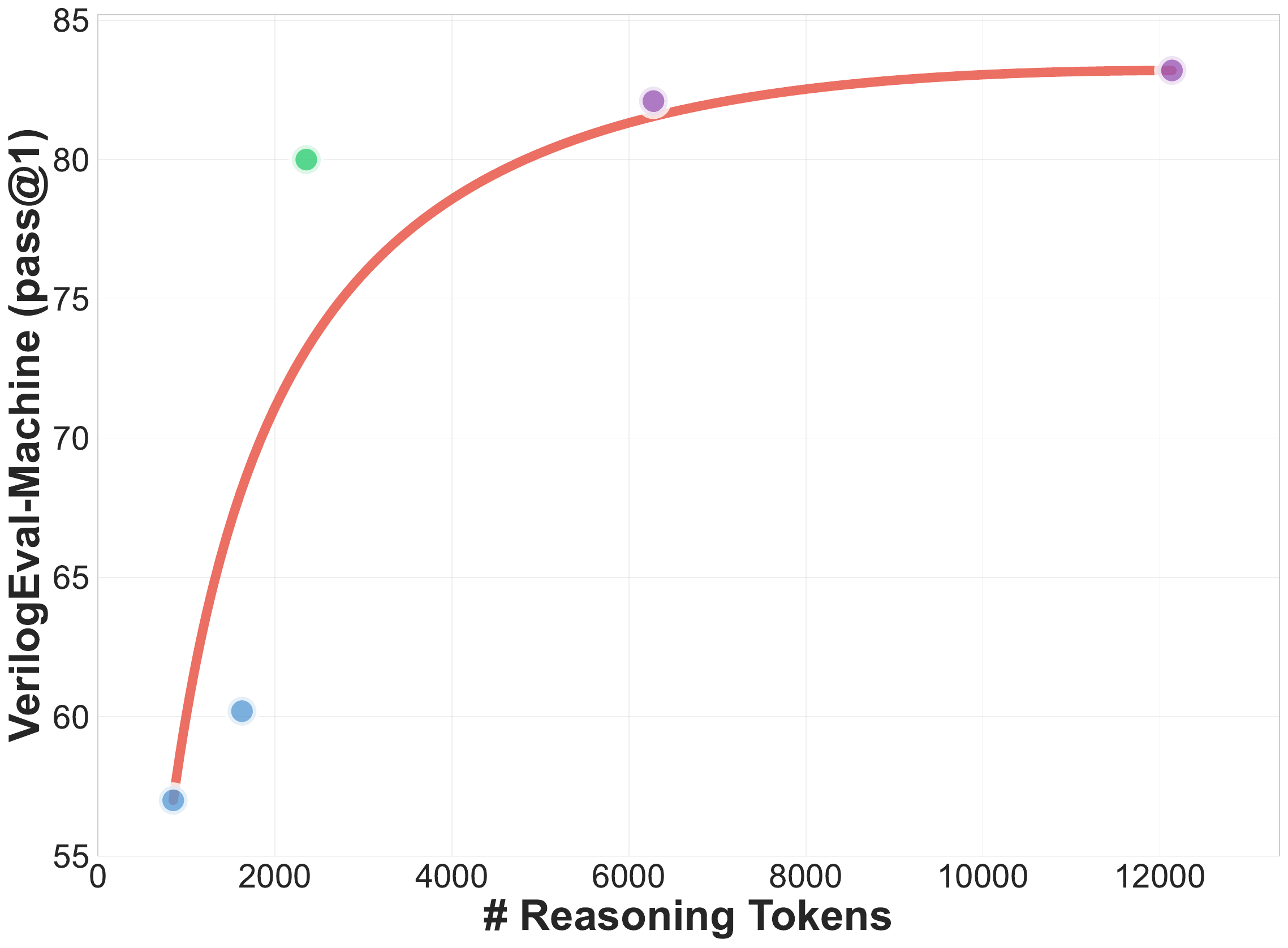}}
    \subfigure[]{%
        \includegraphics[width=0.32\textwidth]{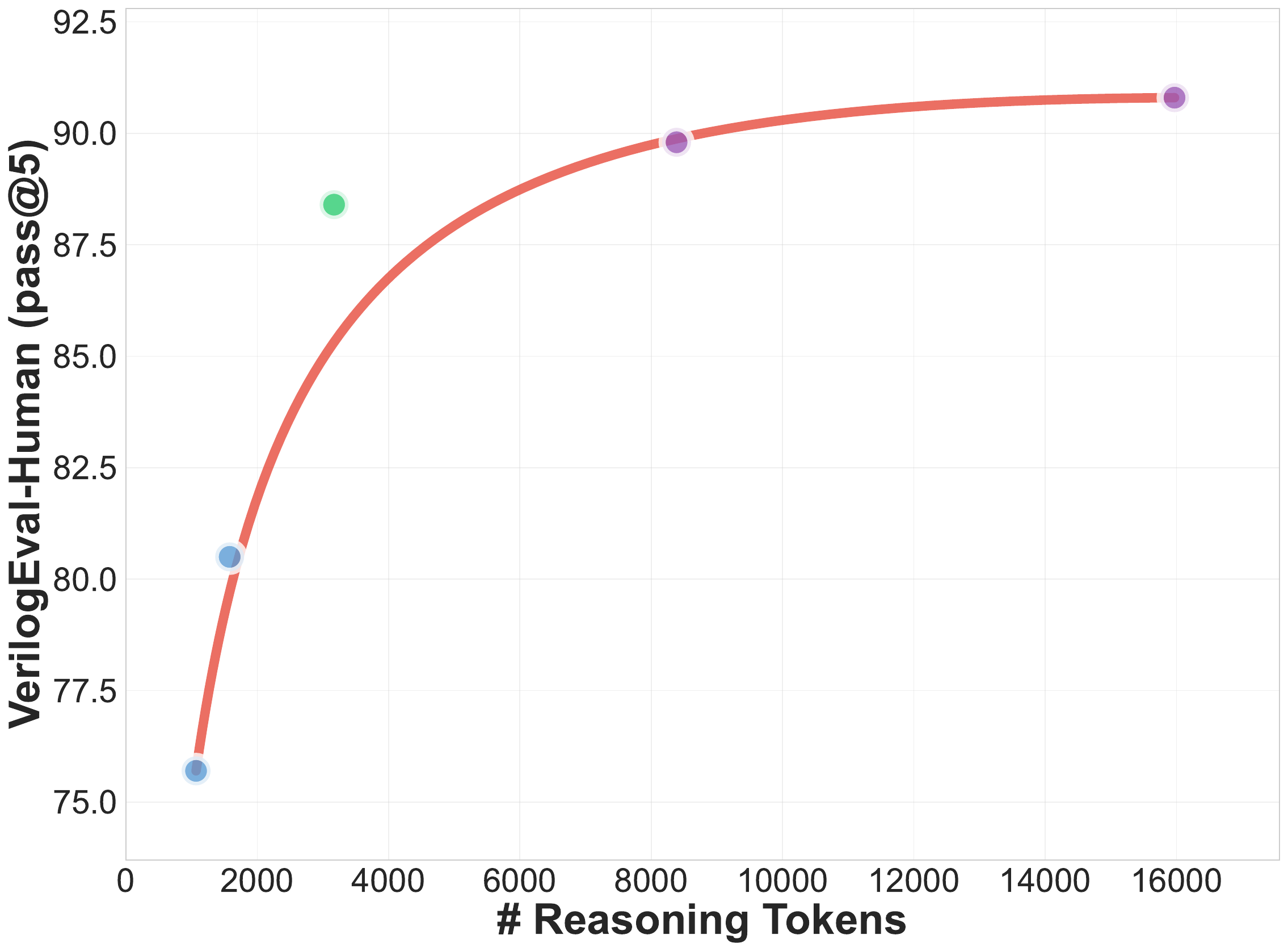}}
    \subfigure[]{%
        \includegraphics[width=0.32\textwidth]{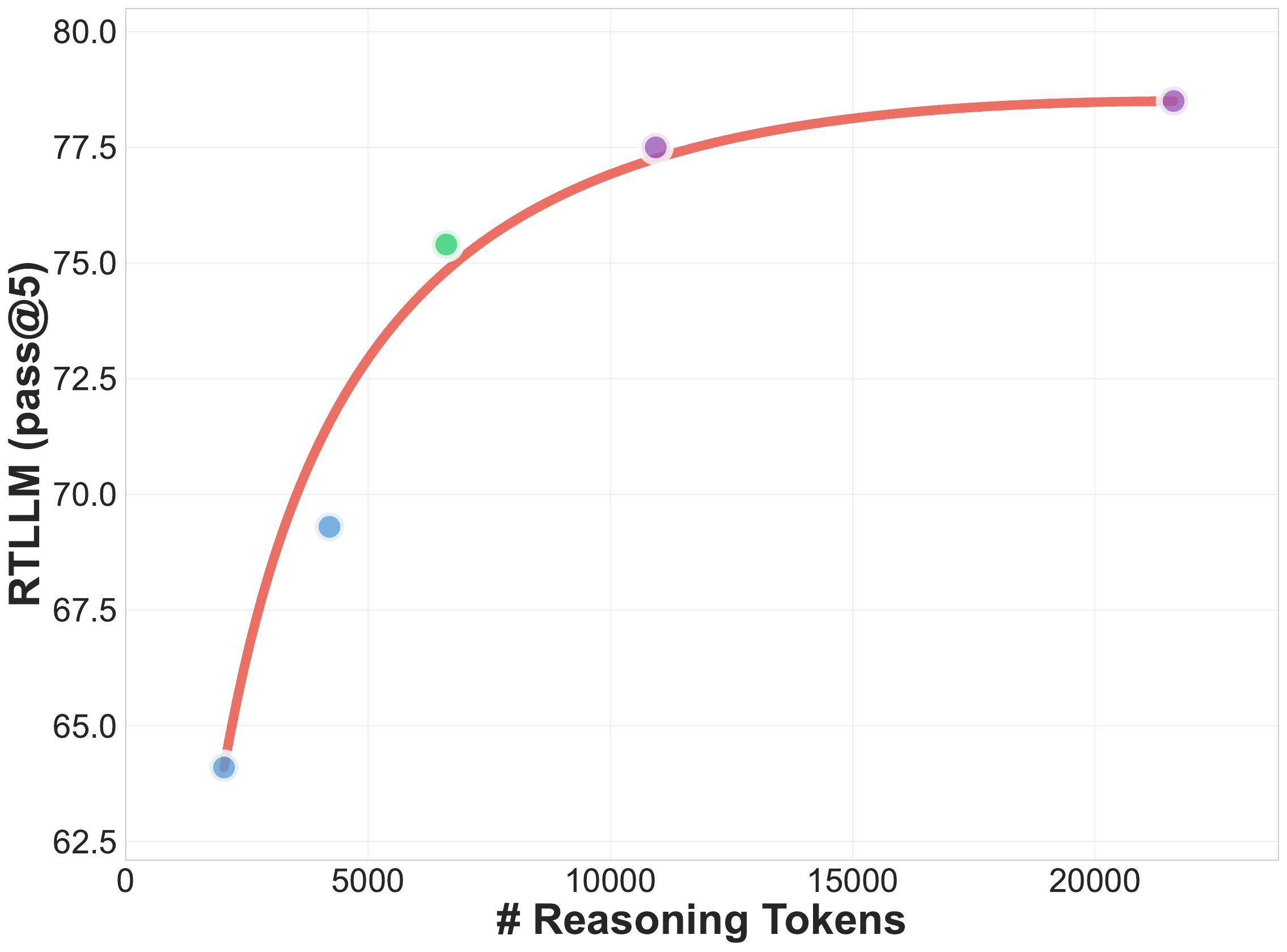}}
    \vspace{-10pt}
    \caption{Impact of reasoning trace length on RTL coding performance --- For each benchmark, we evaluate five \text{\name}$^\dagger$ variants by varying the amount of reasoning. These include two downgraded variants with truncated reasoning traces (\textcolor{NavyBlue}{blue}), the base \name model (\textcolor{Green}{green}), and two enhanced variants with one and two iterations of test-time scaling that progressively extend the reasoning process (\textcolor{Purple}{purple}).}
    \label{fig:rtl_comparison}
\end{figure*}
\begin{figure*}[t!]
\begin{center}
	\includegraphics[width=\textwidth]{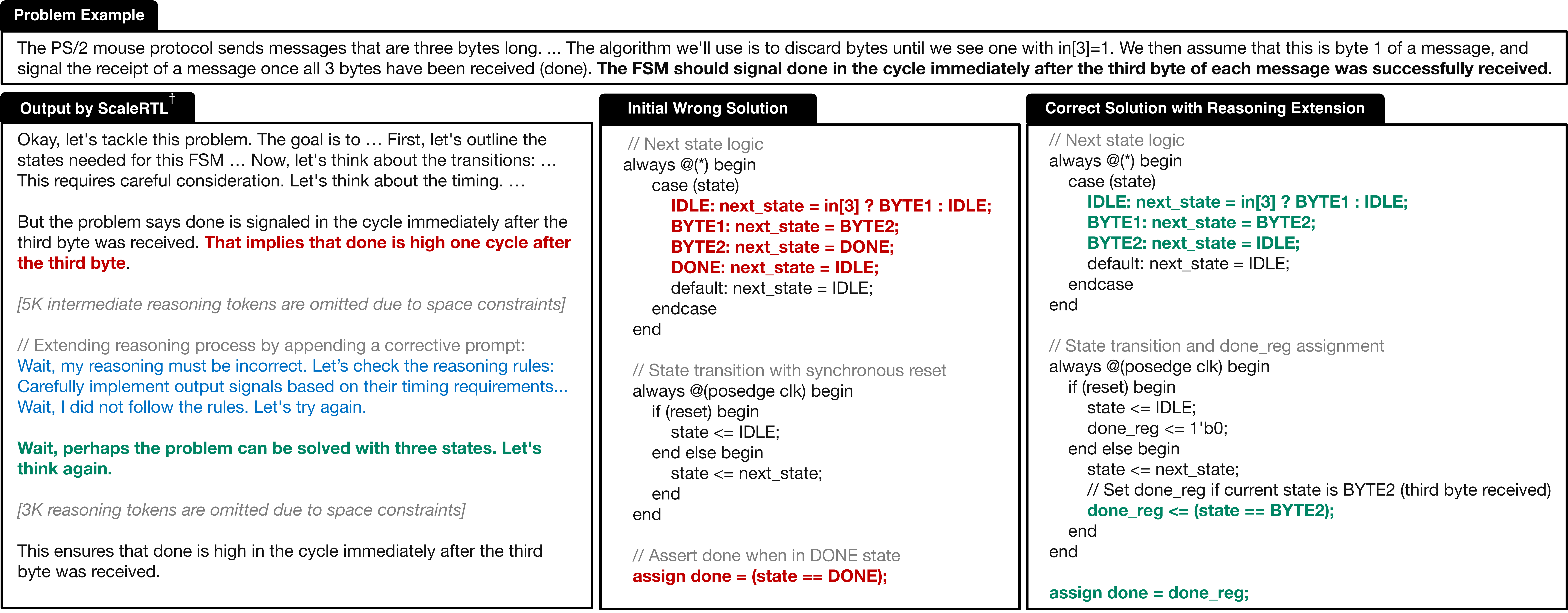}
	\caption{A test-time scaling example in VerilogEval-Human.} 
\vspace{-15pt}
     \label{figure:test_scale}
\end{center}
\end{figure*}
Recent work has empirically observed test-time scaling laws, where model performance improves as more tokens are allocated during inference~\cite{snell2024scaling}. This behavior often follows a log-linear or power-law relationship, particularly in reasoning tasks such as mathematical problem solving. In this section, we investigate whether a similar trend exists in the domain of RTL code generation.

To this end, we analyze the performance of five \name variants across three benchmarks: VerilogEval-Machine, VerilogEval-Human, and RTLLM. These variants include: (1) the base \name, (2) the model after applying one iteration of our test-time scaling approach, and (3) the model after two iterations, which progressively extend the reasoning process. In addition, we generate two downgraded variants by truncating the reasoning trace of \name. Specifically, we suppress its reasoning in the middle by replacing the keyword ``\texttt{Wait}'' with the delimiter ``\texttt{</think>},'' which signals the end of reasoning and forces the model to directly output the RTL solution. For each of these five settings, we compute the average number of reasoning tokens and the corresponding RTL coding performance.

Figures~\ref{fig:rtl_comparison}(a), (b), and (c) present the results for the three benchmarks, respectively. For VerilogEval-Human and RTLLM, we report pass@5 as the evaluation metric, as it demonstrates greater robustness in reflecting model performance. Across all benchmarks, we observe a consistent trend where model accuracy improves rapidly with increased reasoning length and gradually saturates, resembling a logarithmic curve. This highlights the importance of test-time reasoning as a controllable and effective mechanism for enhancing performance without additional training. Furthermore, the smooth curve fit across the five data points in each benchmark indicates that the relationship between reasoning depth and accuracy is systematic and predictable, echoing trends observed in general language reasoning tasks.

To further illustrate the impact of test-time scaling, Figure~\ref{figure:test_scale} presents a concrete example where extending the reasoning process leads to a correct solution. The task involves designing a finite state machine (FSM) that detects valid PS/2 mouse messages, which consist of three bytes. The key challenge lies in correctly implementing the timing behavior: ``\textit{the \texttt{done} signal must be asserted in the same cycle as the third byte is received}''. Initially, \name incorrectly reasons that the \texttt{done} signal should be asserted in the cycle following the receipt of the third byte. This leads to a faulty RTL implementation where \texttt{done} is derived from a registered signal, resulting in a one-cycle delay. After applying our test-time scaling approach by replacing the original end-of-reasoning delimiter \texttt{</think>} with a corrective prompt, the model continues its reasoning, revisits the timing specification, and identifies the flaw in its initial logic. It then generates a correct solution by directly asserting \texttt{done} in the same cycle as the state transition upon receiving the third byte. This example illustrates how extending the reasoning trace enables the model to refine its understanding of precise timing requirements and produce a correct RTL implementation.

Our results confirm that test-time scaling laws hold in the RTL domain. Longer reasoning at inference leads to more accurate RTL code without extra fine-tuning, offering a promising path to improve performance with minimal deployment overhead.

\subsection{General Capability Analysis}
\label{general}
\begin{table}[t]
\centering
\caption{Results on general coding and instruction-following tasks --- We report pass@1 for HumanEval and MBPP, and accuracy for IFEval. We highlight the top \textcolor{Green}{\textbf{first}}, \textcolor{NavyBlue}{\textbf{second}}, and \textcolor{Purple}{\textbf{third}} results per benchmark.}
\label{table:general_benchmarks}
\begin{tabular}{lccc}
\toprule
 & \textbf{HumanEval} & \textbf{MBPP} & \textbf{IFEval} \\
\midrule
DeepSeek-R1-Distill-Qwen-32B & $49.1$ & \textcolor{NavyBlue}{\textbf{71.3}} & \textcolor{Green}{\textbf{76.8}} \\
CodeV-6.7B & \textcolor{Purple}{\textbf{60.2}} & \textcolor{Purple}{\textbf{66.8}} & $24.3$ \\
OriGen-6.7B & \textcolor{NavyBlue}{\textbf{60.7}} & $48.4$ & $46.6$ \\
CraftRTL-15B & $45.7$ & $63.7$ & \textcolor{Purple}{\textbf{39.8}} \\
ScaleRTL-32B & \textcolor{Green}{\textbf{65.2}} & \textcolor{Green}{\textbf{75.0}} & \textcolor{NavyBlue}{\textbf{73.2}} \\
\bottomrule
\vspace{-20pt}
\end{tabular}
\end{table}

Beyond achieving strong performance on RTL benchmarks, it is equally important for RTL coding models to maintain critical capabilities such as solving general programming problems and strictly adhering to user instructions. These abilities reflect fundamental reasoning and alignment skills that are beneficial for complex RTL tasks in practice, which often require careful interpretation of specifications and precise logical reasoning. A model that lacks such capabilities risks overfitting to narrow RTL coding problems in those benchmarks and may fail when applied to more open-ended or unfamiliar design challenges. To evaluate this aspect, we assess \name and prior RTL models on general-purpose coding benchmarks HumanEval and MBPP, as well as the instruction-following benchmark IFEval, which includes 541 prompts designed to test a model's ability to follow diverse and structured instructions.

The comparison shown in Table~\ref{table:general_benchmarks}
reveals a key limitation of existing RTL-specific LLMs: while they achieve promising performance on RTL coding tasks, they fall short on general benchmarks, particularly IFEval, indicating that their specialized training compromises broader core LLM abilities. 
In contrast, \name-32B demonstrates strong general capabilities, achieving results on par with, and in some cases surpassing, the original DeepSeek-R1-Distill-Qwen-32B on all benchmarks. 
This balance of domain specialization and general competency highlights the strength of our reasoning-centric adaptation strategy, which enhances RTL performance without eroding the model’s foundational versatility. This renders \name better suited for practical and robust deployment.


\section{Conclusion}
This work presents the first RTL coding LLM that scales reasoning data and test-time compute. This is acheived by fine-tuning on a large-scale CoT dataset enriched with RTL reasoning patterns. At inference time, we further improve performance through a novel test-time scaling strategy that enables iterative reflection and self-correction. Experimental results showcase that our approach achieves state-of-the-art performance on RTL benchmarks.


\bibliographystyle{plain}
\bibliography{ref}

\clearpage
\appendices
\section{Prompts for Reasoning Data Curation}
\label{prompt_spec}
\begin{figure}[ht]
\begin{tcolorbox}[title=Prompt to generate specification]
\label{figure:spec_prompt}
\footnotesize
Your goal is to create a high-quality Verilog problem.

\begin{itemize}
    \item \textbf{Guidelines for designing the problem description:}
    \begin{enumerate}
        \item This should be \textbf{completely self-contained}, providing all the contextual information one needs to understand and solve the problem.
        \item Assume common Verilog knowledge, but ensure that any specific context, variables, or code snippets pertinent to this problem are explicitly included.
        \item Do not include the code snippet in the problem.
        \item The problem should be designed for the programmer to solve with one Verilog module. Here is an example:
    \end{enumerate}

    \item \textbf{Guidelines for the problem description format:}

    The problem description section should be enclosed within \texttt{<PROBLEM>} \texttt{</PROBLEM>} tags.  
    Below shows an example:

\begin{lstlisting}
Output:
<PROBLEM>
Build a counter that counts from 0 to 999, inclusive, with a period of 1000 cycles.
The reset input is active high synchronous, and should reset the counter to 0.
Solve the problem by completing the following module.
</PROBLEM>
\end{lstlisting}

    \item Now, please gain inspiration from the following code snippet to create a high-quality Verilog problem.
    
    \item Please increase the difficulty of the given programming test question a bit. You can increase the difficulty using, but not limited to, the following methods:
    \begin{enumerate}
        \item Your new problem should not be directly solved by the original code snippet.
        \item You can also change the bit-width requirement, how to reset internal signals (if applicable), and whether the solution needs a clock signal (combinatorial versus sequential logic). If you do have a reset method that is synchronous to a clock, make sure to add the clock signal to the problem module input.
        \item Add new constraints and requirements to the original problem, adding approximately 10 additional words.
        \item Replace a commonly used requirement in the programming task with a less common and more specific one.
        \item If the original problem can be solved with only a few logical steps, please add more reasoning steps.
    \end{enumerate}
\end{itemize}

\textbf{Code snippet for inspiration:}
\begin{verbatim}
{code}
\end{verbatim}

\noindent\textbf{Output:}

\end{tcolorbox}
\caption{Prompt for specification generation}
\end{figure}
\begin{figure}[t!]
\begin{tcolorbox}[title=Prompt to generate Verilog solution with reasoning]
\footnotesize
Your goal is to solve a Verilog problem by completing one Verilog module.

\begin{itemize}
\item \textbf{Guidelines for writing the solution (Verilog module) and formatting:}
\begin{enumerate}
    \item The module should be \textbf{completely self-contained}, fulfilling all the requirements needed to solve the problem.
    \item Ensure the correctness of the syntax and functionality.
    \item The solution must be valid Verilog code.
    \item Do not include or import outside the module and define everything inside the module.
    \item The solution section should be enclosed within \texttt{<answer>} \texttt{</answer>} tags.
\end{enumerate}

\item \textbf{Below shows an example:}

\textit{Problem description:}
\begin{lstlisting}[basicstyle=\ttfamily\footnotesize, breaklines=true]
Build a counter that counts from 0 to 999, inclusive, with a period of 1000 cycles. 
...
\end{lstlisting}

\textit{Output:}

<think>
\begin{lstlisting}[basicstyle=\ttfamily\footnotesize, breaklines=true]
Okay, so I need to design a counter that goes from 0 to 999 and then repeats. The module is called top_module and has clk, reset, and output q which is a 10-bit register.
...

Yes. The code should handle all that.
\end{lstlisting}

</think>

<answer>
\begin{lstlisting}[language=verilog]
module top_module(
    input clk,
    input reset,
    output reg [9:0] q);

    always @(posedge clk) begin
        if (reset) begin
            q <= 10'd0; // Synchronous reset to 0
        end else begin
            if (q == 10'd999) begin // Check if reached 999
                q <= 10'd0;
            end else begin
                q <= q + 10'd1; // Increment otherwise
            end
        end
    end

endmodule

\end{lstlisting}
</answer>

\item Now, please solve the following Verilog problem. I will also attach a reference code snippet which was used as an inspiration to generate the problem. The provided code may not directly solve the problem, so use it only as a reference.

\textbf{Reference code:}
\begin{lstlisting}[basicstyle=\ttfamily\footnotesize, breaklines=true]
{reference_code}
\end{lstlisting}

\textbf{Problem description:}
\begin{lstlisting}
{problem}
\end{lstlisting}

\textbf{Output:}
\end{itemize}
\end{tcolorbox}
\caption{Prompt for generating Verilog solution with reasoning}
\end{figure}


\clearpage

\section{Prompt for Reasoning Rule Generation}
\label{prompt_rule}
\begin{figure}[ht]
\begin{tcolorbox}[title=Prompt to generate general RTL rules]
\footnotesize
You are an expert in Verilog hardware description language, focusing on analyzing code generation challenges and potential pitfalls.

\vspace{1em}
I will provide you with a markdown file containing a Verilog code generation benchmark, which includes:
\begin{itemize}
    \item The original prompt/problem statement
    \item Additional context and information
\end{itemize}

\vspace{1em}
\textbf{\# Your Task} \\
Analyze the Verilog code generation task in depth and provide:

\begin{enumerate}
    \item \textbf{Potential Pitfalls Analysis}: Identify the key challenges and tricky aspects of this problem that could lead to incorrect code generation. This could include:
    \begin{itemize}
        \item Complex syntax requirements
        \item Subtle logic implementation details
        \item Problem interpretation challenges
        \item Timing-sensitive operations
        \item Signal width considerations
        \item Tricky Verilog construct usage
    \end{itemize}

    \item \textbf{Critical Implementation Areas}: Point out the specific aspects of the problem that are most prone to errors, with clear explanations of what makes these areas challenging.

    \item \textbf{Problem Complexity Assessment}: Analyze the inherent complexity of the problem and identify areas where careful attention to detail is required for correct implementation.

    \item \textbf{General Verilog Coding Rules}: Based on the potential challenges identified, formulate concise, general Verilog coding rules that would help prevent common mistakes. These rules should:
    \begin{itemize}
        \item Be widely applicable to Verilog coding, not just this specific example
        \item Focus on best practices for correctness, not just style
        \item Be clear and actionable
        \item Address the potential pitfalls you identified
        \item Focus \textbf{only} on rules that help generate syntactically and functionally correct RTL code
    \end{itemize}
\end{enumerate}

\vspace{1em}
Format your response as follows:

\vspace{1em}
\textbf{Potential Pitfalls Analysis} \par
[Your detailed analysis of challenging aspects that could lead to incorrect code generation]

\textbf{Critical Implementation Areas} \par
[Your identification of error-prone areas in the implementation]

\textbf{Problem Complexity Assessment} \par
[Your analysis of the problem's inherent complexity and attention-requiring areas]

\textbf{Verilog Coding Rules} \par
[List up to 3 critical Verilog coding rules here, focusing only on correct RTL generation]

\vspace{1em}
Here is the markdown file containing the Verilog code generation benchmark:

\texttt{\{md\_content\}}

\end{tcolorbox}
\caption{Prompt for general RTL rules generation}
\label{figure:rule_prompt}
\end{figure}
To enable \name to rethink and self-correct earlier reasoning steps, we find that incorporating general RTL coding rules into the corrective prompt serves as effective hints, encouraging the model to generate longer chain-of-thought reasoning traces during test-time scaling. We develop an automated method to generate these general rules by prompting Claude3.7-Sonnet with the text shown in Figure \ref{figure:rule_prompt}. This process can be performed once offline and the rules are reused during test-time scaling. Notably, hardware engineers can incorporate RTL rules based on their own domain knowledge in practice, which may further enhance the model’s reasoning process.

\section{Comparison against SoTA Foundation LLMs}
\label{sota_llms}
\begin{table}[h]
\centering
\caption{Results on RTL coding benchmarks --- We report pass@1 on VerilogEval-Machine (VerilogEval-M), VerilogEval-Human (VerilogEval-H), and pass@5 on RTLLM v1.1. ScaleRTL-32B and its test-time scaling variant (\text{\name}$^\dagger$-32B) are compared against strong proprietary foundation models. We highlight the top \textcolor{Green}{\textbf{first}}, \textcolor{NavyBlue}{\textbf{second}}, and \textcolor{Purple}{\textbf{third}} results per benchmark.}
\label{table:sota_llms}
\begin{adjustbox}{width=\columnwidth,center}
\begin{tabular}{lccc}
\toprule
 & \textbf{VerilogEval-M} & \textbf{VerilogEval-H} & \textbf{RTLLM v1.1} \\
\midrule
Claude3.5-Sonnet & \textcolor{Green}{\textbf{83.4}} & $73.7$ & $44.8$ \\
Claude3.7-Sonnet & $70.0$ & $73.0$ & $48.2$ \\
GPT4.1 & $72.9$ & $68.6$ & $44.8$ \\
OpenAI o3 & $77.0$ & \textcolor{NavyBlue}{\textbf{80.0}} & $54.5$ \\
OpenAI o4-mini & $73.2$ & $77.3$ & $44.8$ \\
DeepSeek-R1 & $79.7$ & \textcolor{Purple}{\textbf{79.3}} & \textcolor{Purple}{\textbf{56.7}} \\
\midrule
ScaleRTL-32B & \textcolor{Purple}{\textbf{80.0}} & $76.3$ & \textcolor{NavyBlue}{\textbf{75.4}} \\
\text{\name}$^\dagger$-32B & \textcolor{NavyBlue}{\textbf{83.2}} & \textcolor{Green}{\textbf{80.4}} & \textcolor{Green}{\textbf{78.5}} \\
\bottomrule
\end{tabular}
\end{adjustbox}
\end{table}

To further validate the efficacy of our approach, we compare \name-32B and its test-time scaling variant \text{\name}$^\dagger$-32B against strongest proprietary foundation models, including Claude3.5-Sonnet, Claude3.7-Sonnet, GPT-4.1, OpenAI o3, o4-mini, and DeepSeek-R1. It is important to note that the training data cutoff dates for these proprietary models are after the public release of both VerilogEval and RTLLM benchmarks. As a result, it is highly likely that their training corpora contain benchmark-related information, which may lead to overestimated performance on these evaluation sets.

In contrast, we have carefully filtered our training data to ensure no information from VerilogEval or RTLLM benchmarks was included during fine-tuning. Despite this more rigorous setup, our best-performing model, \text{\name}$^\dagger$-32B, achieves state-of-the-art results across all three RTL benchmarks. It outperforms all baselines on VerilogEval-Human and RTLLM, achieving $80.4\%$ and $78.5\%$, respectively. Additionally, \text{\name}$^\dagger$-32B attains $83.2\%$ on VerilogEval-Machine, closely matching or exceeding the performance of the strongest closed-source models. These results highlight the efficacy of our approach, even without access to potentially leaked benchmark data.

\section{Pass Rate Evaluation on RTL Benchmarks}
\label{pass_rate}
\begin{table}[h]
\centering
\caption{Pass rate on VerilogEval-Machine (VerilogEval-M), VerilogEval-Human (VerilogEval-H), and RTLLM v1.1.}
\label{table:pass_rate}
\begin{adjustbox}{width=\columnwidth,center}
\begin{tabular}{lccc}
\toprule
 & \textbf{VerilogEval-M} & \textbf{VerilogEval-H} & \textbf{RTLLM v1.1} \\
\midrule
ScaleRTL-32B & $88.1$ & $90.4$ & $79.3$ \\
\text{\name}$^\dagger$-32B & $\mathbf{90.2}$ & $\mathbf{93.6}$ & $\mathbf{82.8}$ \\
\bottomrule
\end{tabular}
\end{adjustbox}
\end{table}

Unlike pass@k, which averages over multiple completions, pass rate is defined as the percentage of problems for which the model succeeds at least once in $n$ trials ($n=10$ in this work). It more directly captures the effectiveness of our iterative reasoning approach, where each inference attempt builds on extended reasoning traces.Table~\ref{table:pass_rate} shows that \text{\name}$^\dagger$-32B consistently outperforms \name-32B across all benchmarks, demonstrating that longer reasoning at test time is effective for accurate RTL code generation.

\end{document}